\documentclass[aps, prd, amsmath, floats, floatfix, twocolumn, nofootinbib, superscriptaddress, showpacs]{revtex4-1}
\usepackage{graphicx,color}
\usepackage[dvipsnames]{xcolor}  
\usepackage{amsmath,amsfonts,amssymb}
\usepackage{physics}     
\usepackage{braket}
\usepackage{dsfont}      
\usepackage{varwidth}
\usepackage{hyperref}
\usepackage[utf8]{inputenc}

\def\bR{{\mathbb R}}

\def\bHo{{\mathring{\mathbb H}}}

\def\PAdS{\text{PAdS}}

\begin{document}


\title{Mode solutions for a Klein-Gordon field in anti-de Sitter spacetime \\ with dynamical boundary conditions of Wentzell type}


\author{Claudio Dappiaggi}
\email{claudio.dappiaggi@unipv.it}
\affiliation{Dipartimento di Fisica, Universit\`a degli Studi di Pavia, Via Bassi, 6, 27100 Pavia, Italy}

\author{Hugo R. C. Ferreira}
\email{hugo.ferreira@pv.infn.it}
\affiliation{Istituto Nazionale di Fisica Nucleare -- Sezione di Pavia, Via Bassi, 6, 27100 Pavia, Italy}

\author{Benito A. Ju\'arez-Aubry}
\email{benito.juarez@correo.nucleares.unam.mx}
\affiliation{Departamento de Gravitaci\'on y Teor\'ia de Campos, Instituto de Ciencias Nucleares, Universidad Nacional Aut\'onoma de M\'exico, A. Postal 70-543, Mexico City 045010, Mexico.}


\date{\today}

\begin{abstract}
We study a real, massive Klein-Gordon field in the Poincar\'e fundamental domain of the $(d+1)$-dimensional anti-de Sitter (AdS) spacetime, subject to a particular choice of \emph{dynamical} boundary conditions of generalized Wentzell type, whereby the boundary data solves a non-homogeneous, boundary Klein-Gordon equation, with the source term fixed by the normal derivative of the scalar field at the boundary. This naturally defines a field in the conformal boundary of the Poincar\'e fundamental domain of AdS. We completely solve the equations for the bulk and boundary fields and investigate the existence of bound state solutions, motivated by the analogous problem with Robin boundary conditions, which are recovered as a limiting case. Finally, we argue that both Robin and generalized Wentzell boundary conditions are distinguished in the sense that they are invariant under the action of the isometry group of the AdS conformal boundary, a condition which ensures in addition that the total flux of energy across the boundary vanishes.
\end{abstract}

\maketitle

\section{Introduction}

Classical and quantum field theory on asymptotically anti-de Sitter (AdS) spacetimes, and generally other spacetimes with boundaries, has been the target of significant attention in the last two decades, mainly inspired by the remarkable AdS/CFT correspondence \cite{Maldacena:1997re, Witten:1998qj}, see \cite{Ammon:2015} for a modern overview. The importance of this correspondence has gone well beyond its initial connection with the quantum gravity formulation in string theory and has become relevant in many low energy physics applications, ranging from nuclear to condensed matter physics \cite{Hartnoll:2009sz}.

From a geometric standpoint, in contrast with their asymptotically flat or asymptotically de Sitter counterparts, asymptotically AdS spacetimes are not globally hyperbolic; the conformal asymptotic boundary at infinity is timelike. As a consequence, on an asymptotically AdS background, one cannot expect to find global solutions for hyperbolic equations, such as the Klein-Gordon one, only by assigning suitable initial data. These must be supplemented with appropriate boundary conditions imposed at the conformal boundary. 

In previous work \cite{Dappiaggi:2016fwc,Dappiaggi:2017wvj}, two of the authors analyzed the classical and quantum field theory of a massive scalar field propagating in anti-de Sitter (AdS) spacetime in $d+1$ spacetime dimensions, subject to homogeneous Robin boundary conditions, which include the familiar Dirichlet and Neumann boundary conditions as particular cases, see also \cite{Bussola:2017wki} for an analysis on BTZ spacetime. In that work, by means of a Fourier transform, the Klein-Gordon equation has been reduced to a Sturm-Liouville problem, which naturally provides all the admissible boundary conditions of Robin type for a specific range of the mass parameter of the field. In this context, studying all admissible Robin boundary conditions at once is a good strategy for finding the parameter space in mass and curvature coupling for which there exist bound state solutions, which decay exponentially away from the AdS boundary. These modes not only lead to instabilities in the classical linear theory but also pose an obstruction to the existence of a ground state for the underlying quantum theory.

These results call for two natural generalizations, the first by allowing the background to be a generic asymptotically AdS spacetimes, the second by fixing more general boundary conditions. In particular, in the AdS case, the second avenue implies in particular that one should treat boundary value problems outside of the realm of Sturm-Liouville theory.

From a structural standpoint, since there exist infinite choices of boundary conditions, the first step consists of identifying a natural subclass which is distinguished for its physical properties. To this end, it seems that one bit of information that could be used is the existence of a large group of isometries at the conformal boundary. Heuristically one expects that boundary conditions should be chosen in such a way to be compatible with the action of such group. The problem of translating such idea in a concrete mathematical tool can be addressed in the specific case of an AdS spacetime by adapting and reinterpreting the recent results of \cite{Ibort:2014sua, Ibort2, Pardo}. In particular one can realize that each boundary condition is nothing but an operator acting between (a suitable generalization of) the field restricted to the boundary and its normal derivative. From this perspective it is natural to require that such operator commutes with the scalar representation of the isometry group of the conformal boundary. By using this paradigm one restricts considerably the class of possible boundary conditions, while, at the same time, making clear that it is possible to go beyond those of Robin type.

In this paper, we study a massive scalar field in AdS in $d+1$ spacetime dimensions, AdS$_{d+1}$, subject to {\it generalized Wentzell boundary conditions} (WBCs). Specifically, we focus on the so-called Poincar\'e fundamental domain PAdS$_{d+1}$, which covers only a portion of the full AdS$_{d+1}$ and it has the advantage of being conformal to the half-Minkowski spacetime in $d+1$ dimensions, $\bHo^{d+1}$. As we shall see below, these boundary conditions have boundary data determined by a non-homogeneous, boundary Klein-Gordon equation, with the source term fixed by the normal derivative of the scalar field at the boundary. This naturally defines a field in the conformal boundary of the Poincar\'e fundamental domain of AdS, a feature which is clearly reminiscent of the AdS/CFT framework, though here we limit ourselves to considering non interacting models. 

In our investigation, although we limit ourselves to considering classical features of the underlying model, the existence of bound states in particular, our ultimate goal is to provide a full-fledged quantum system. This is in particular one of the key reasons why we shall not only focus on finding smooth solutions for the underlying dynamics, but we will also be interested in the square-integrability of the relevant functions.

As we show in this paper, the WBCs are \emph{distinguished} in the sense that they are dynamical boundary conditions which are invariant under the action of the isometry group of the AdS boundary. Furthermore, as is the case with the simpler, nondynamical Robin boundary conditions, the total fluxes of symplectic and energy currents across the AdS conformal boundary vanish, as required for a closed system. Moreover, the Robin boundary condition eigenstate solutions to the Klein-Gordon operator can be recovered as suitable limits of the WBC ones.

The treatement of WBC boundary conditions in the classical and quantum field theoretic literature has appeared in the work of one of us, together with Barbero, Margalef-Bentabol and Villase\~nor, \cite{G.:2015yxa, Barbero:2017kvi}, where the simple $(1+1)$-dimensional mechanical model of a finite string with point masses subject to harmonic potentials in the extrema has been studied in detail. In that work, the classical system is solved and the Fock quantization is performed, with the ultimate goal of constructing {\it boundary Hilbert spaces} where the dynamics of the extremal masses takes place, with the aid of the PDE {\it Lions trace} operators. It is further shown that the quantum mechanical dynamics in the boundary Hilbert spaces is non-unitary. 

These boundary conditions have also been considered in \cite{Zahn:2015due} in $(d+1)$-dimensional Minkowski spacetime with one or two timelike boundaries. There, it was investigated the Fock space quantization of the underlying system and, in addition it was shown that the WBC ensure that the short-distance singularities of the two-point function for the boundary field has the form expected of a field living in a $d$-dimensional spacetime, contrary to other boundary conditions, for which the two-point function inherits the short-singularity of the $(d+1)$-dimensional bulk. This seems to be a very desired feature for holographic purposes. Previous explorations of the WBC in mathematical literature appear in \textit{e.g.}~\cite{Ueno:1973,Favini:2002,Coclite:2014}.

While the main inspiration of this work comes from high energy physics, namely the AdS/CFT conjecture, and the connection to holographic renormalization \cite{Skenderis:2002wp} is a central motivation for us, we note that the boundary conditions that we consider, as well as related dynamical boundary conditions, are suitable for studying systems in a broad sprectrum of physical problems. We point out that dynamical boundary conditions are generally interesting from the point of view of modelling open systems in condensed matter physics. They are also relevant for lower-dimensional Chern-Simons theories coupled to electrodynamics, which can model e.g. effective topological insulators \cite{Martin-Ruiz:2015skg}. From a gravitational perspective, dynamical boundary conditions are interesting in the study of isolated horizons, providing an avenue for associating degrees of freedom at a horizon surface. This is attractive from a quantum gravity perspective. In loop quantum gravity, for example, a procedure for counting horizon degrees of freedom yields the Bekenstein-Hawking entropy \cite{Ashtekar:1997yu, Ashtekar:1999wa}.

The contents of this paper are as follows. In Sec.~\ref{sec:AdS} we review the basic geometric properties of AdS$_{d+1}$ and the closely related PAdS$_{d+1}$, the Poincar\'e fundamental domain of AdS$_{d+1}$ as a spacetime in its own right. In addition we show how the Klein-Gordon equation on PAdS$_{d+1}$ can be treated as a (generally singular) Klein-Gordon equation on half-Minkowski. In Sec.~\ref{sec:WBCs} we introduce WBCs. We motivate them in the form of an action principle for the Klein-Gordon field with boundary dynamical contributions in half-Minkowski spacetime. We then deal with the problem in PAdS$_{d+1}$ with the aid of the aforementioned conformal techniques, by solving both the bulk and boundary field equations in full generality. We consider separately two cases: in Sec.~\ref{sec:regular} the regular case, corresponding to the massless, conformally coupled (conformally transformed) scalar field (in half-Minkowski), and in Sec.~\ref{sec:singular} the singular case, corresponding to the general, massive scalar field. Additionally, in both cases, we investigate the existence of bound state mode solutions, which decay exponentially away from the boundary. Finally, in Sec.~\ref{sec:maths}, we explicitly show the vanishing of the fluxes of symplectic and energy currents across the AdS boundary when WBCs are imposed, and explain how these boundary conditions are invariant under the boundary isometry group, making them distinguished.

Throughout the paper we employ natural units in which $c = G_{\rm N} = 1$ and a metric with signature $({-}{+}{+}\cdots)$.

\section{Anti-de Sitter spacetime and Klein-Gordon field}
\label{sec:AdS}

In this section, we briefly review the basic geometric properties of anti-de Sitter spacetime AdS$_{d+1}$ and introduce the Poincar\'e fundamental domain PAdS$_{d+1}$, on which we consider a classical scalar field satisfying the Klein-Gordon equation.

\subsection{Geometry of anti-de Sitter spacetime}

The maximally symmmetric solution of the Einstein field equations with negative cosmological constant, $\Lambda$, is the anti-de Sitter spacetime, which we denote by AdS$_{d+1}$ in $d+1$ Lorentzian dimensions. As a manifold it is diffeomorphic to $\mathbb{S}^1 \times \mathbb{R}^d$ and it can be realized as an embedded submanifold in the ambient space $\mathbb{R}^{d+2}$ equipped with the metric
\begin{equation}
g_{\mathbb{R}^{d+2}} = - \dd X_0^2 - \dd X_1^2 + \sum_{i,j = 2}^{d+1} \delta^{ij} \, \dd X_i \dd X_j \ ,
\label{Rd2metric} 
\end{equation} 
via the equation
\begin{equation}
-X_0^2 - X_1^2 + \sum_{i = 2}^{d+1} X_i^2 = \frac{d(d-1)}{\Lambda}\,
\end{equation}
where each $X_i$, $i=0,...,d+2$ is a Cartesian coordinate, while $\delta^{ij}$ stands for the Kronecker delta. As a consequence AdS$_{d+1}$ comes equipped with the induced (Lorentzian, non-degenerate) metric. 

We can give an explicit representation of these geometric structures by considering the {\it Poincar\'e fundamental domain} of AdS$_{d+1}$, denoted by PAdS$_{d+1}$. This is covered by the {\it Poincar\'e coordinate patch}, with $t \in \mathbb{R}$, $z \in \mathbb{R}^+$ and $x_i \in \mathbb{R}$, defined by 
\begin{align}
\left\{
                \begin{array}{ll}
                  X_0 = \cfrac{\ell}{z} t, \\
                  \vspace*{-0.2cm} \\                  
                  X_1 = \cfrac{z}{2} \left(1 + \cfrac{1}{z^2} \left(- t^2 + \displaystyle\sum_{i = 1}^{d-1} x_i^2 + \ell^2 \right) \right), \\
                  \vspace*{-0.2cm} \\
                  X_i = \cfrac{\ell}{z} x_{i-1}, \quad i = 2, \ldots, d,\\
				  \vspace*{-0.2cm} \\                  
                  X_{d+1} = \cfrac{z}{2} \left(1 + \cfrac{1}{z^2} \left(- t^2 + \displaystyle\sum_{i = 1}^{d-1} x_i^2 - \ell^2 \right) \right),
                \end{array}
              \right.
\label{PAdScoods}
\end{align}
where $\ell^2 = -d(d-1)/ \Lambda$. Thus, PAdS$_{d+1}$ is a Lorentzian spacetime with underlying manifold $\mathbb{R}^+ \times \mathbb{R}^d$ equipped with the metric
\begin{equation}
g_{{\rm PAdS}_{d+1}} = \frac{\ell^2}{z^2} \left( - \dd t^2 + \dd z^2 + \sum_{i, j = 1}^{d-1} \delta^{ij} \dd x_i \dd x_j \right).
\label{gPAdS} 
\end{equation}
Eq.~\eqref{gPAdS} shows that (PAdS$_{d+1},g_{{\rm PAdS}_{d+1}}$) is conformal to the interior of the $(d+1)$-dimensional half-Minkowski spacetime, $(\bHo^{d+1},\eta_{d+1})$, with $\eta_{d+1} = \Omega^2 g_{{\rm PAdS}_{d+1}}$, where the conformal factor is $\Omega = z/\ell$. The conformal boundary of PAdS$_{d+1}$ can be attached at $z = 0$.

\subsection{The Klein-Gordon field in PAdS$_{d+1}$}

In this work, we consider a classical, real Klein-Gordon field $\phi: {\rm PAdS}_{d+1} \to \mathbb{R}$. Given initial data on a hypersurface of PAdS$_{d+1}$ for the Klein-Gordon wave equation,
\begin{equation}
P \phi = \left( \Box_g^{(d+1)} - m_0^2 - \xi R \right) \phi = 0 \ ,
\label{KGPAdS}
\end{equation}
where $\Box_g^{(d+1)}$ is the d'Alembert wave operator, $m_0$ is the mass, $\xi \in \mathbb{R}$ is the coupling to the scalar curvature, while $R = -d(d+1) / \ell^2$ is the Ricci scalar of the spacetime. From now on, we set $\ell = 1$. Although, for initial data, which are smooth and compactly supported in PAdS$_{d+1}$, a unique solution exists in its domain of dependence, in order to address the problem of global existence, one needs to impose boundary conditions at timelike infinity, which, in the Poincar\'e patch, corresponds to the conformal boundary. 

In order to control such freedom, we follow the same strategy adopted in \cite{Dappiaggi:2016fwc} to switch from Eq.~\eqref{KGPAdS} to the associated, conformally-transformed scalar field equation in $\mathbb{\mathring{\mathbb H}}^{d+1}$.  Hence, defining $\Phi = \Omega^{\frac{1-d}{2}}\phi :\bHo^{d+1} \to \mathbb{R}$, the solutions of \eqref{KGPAdS} are in one to one correspondence with those of
\begin{equation}
P_\eta \Phi \doteq \left( \Box_\eta^{(d+1)} - \frac{m^2}{z^2} \right) \Phi = 0 \ ,
\label{KGH}
\end{equation}
where we have defined $m^2 \doteq m_0^2 + (\xi -\frac{d-1}{4d}) R$. A strategy
for dealing with Eq. \eqref{KGPAdS} in the case of a minimally-coupled,
real scalar field in $d = 3$ that does not rely on conformal
transformations has appeared in \cite{Ayon-Beato:2018hxz}. Here, we consider problems for which $m^2 \geq -\frac{1}{4}$, which corresponds to the Breitenlohner-Freedman (BF) bound \cite{Breitenlohner:1982jf}.

The boundary condition imposed at the conformal boundary of PAdS$_{d+1}$ that allows one to obtain global solutions for $\phi$ corresponds in \eqref{KGH} to a boundary condition at $z = 0$, where the potential term becomes singular. In \cite{Dappiaggi:2016fwc, Dappiaggi:2017wvj} all possible homogeneous boundary conditions of Robin type have been analysed. It is the approach of this paper to consider more general, dynamical boundary conditions, which reduce to those of Robin type in a precise limiting sense. In particular, the boundary conditions that we consider are of {\it generalized Wentzell type}. As mentioned above, these have been considered in the mathematical physics literature in \cite{Zahn:2015due} for regular problems in the half-Minkowski spacetime, and also studied by one of the authors in \cite{G.:2015yxa, Barbero:2017kvi} in the context of the quantization of a finite string coupled to point masses subject to harmonic restoring forces at the boundaries.

\section{Wentzell boundary conditions}
\label{sec:WBCs}

In this section, we study the problem defined by Eq.~\eqref{KGH} in the bulk and Wentzell boundary conditions at $z = 0$. In Section \ref{sec:action} we show that these boundary conditions can be obtained naturally starting from an action functional in the so-called regular case ($m^2 = 0$ in \eqref{KGH}). We study such case via a mode expansion in Section~\ref{sec:regular}, finding the conditions under which there exists, together with the expected continuous spectrum, point spectrum contributions to the solutions. These indicate the existence of {\it bound state mode solutions} to the problem. Afterwards we show how to recover the solutions to the problem \eqref{KGH} with Robin boundary conditions, obtaining full agreement with the results reported in \cite{Dappiaggi:2016fwc, Dappiaggi:2017wvj}. In Section~\ref{sec:singular} we repeat the analysis for the singular problem ($m^2 \in [-\frac{1}{4}, \frac{3}{4}) \setminus \{0\}$), also characterizing the spectrum and obtaining the Robin boundary problem solutions in a suitable limit in agreement with \cite{Dappiaggi:2016fwc, Dappiaggi:2017wvj}.

\subsection{Action }
\label{sec:action}

Let us motivate the introduction of the generalized Wentzell boundary conditions, by considering the usual action for a massless Klein-Gordon field in $\bHo^{d+1}$ together with a particular choice of boundary terms:
\begin{align}
S & =  - \frac{1}{2} \int_{t_1}^{t_2} \!\!\!\! \dd t \int_{\mathbb{R}^+} \!\!\!\! \dd z \int_{\mathbb{R}^{d-1}} \!\!\!\!\!\!\!\! \dd^{d-1} x \ \partial_\mu \Phi \partial^\mu \Phi  \nonumber \\
&\quad +\frac{c}{2} \int_{t_1}^{t_2} \!\!\!\! \dd t  \int_{\mathbb{R}^{d-1}} \!\!\!\!\!\!\!\! \dd^{d-1} x \left(-\dot{\Phi}^2 + \partial_i \Phi \partial^i \Phi + m^2_{\rm b} \Phi^2 \right) \, ,
\end{align}
where $c$ and $m_{\rm b}^2$ are arbitrary real parameters at this stage and where repeated indexes are summed over with $\mu \in \{t, z, x_1, \ldots, x_{d-1}\}$ and $i \in \{x_1, \ldots, x_{d-1}\}$, and $\dd^{d-1} x = \prod_{i = 1}^{d-1} \dd x_i$. With a slight abuse of notation, we use the symbol $\Phi$ in the boundary integrand, in place of its restriction thereon. Implicitly we are also restricting our attention to kinematic configurations which are sufficiently regular at $z=0$, to make these operations meaningful.

The variation of the action yields
\begin{align*}
& dS(\Phi) \cdot \delta  = \left. \frac{d}{d \lambda} S(\Phi + \lambda \delta) \right|_{\lambda = 0} \nonumber \\
& = -  \int_{t_1}^{t_2} \!\!\!\! \dd t \int_{\mathbb{R}^+} \!\!\!\! \dd z \int_{\mathbb{R}^{d-1}} \!\!\!\!\!\!\!\! \dd^{d-1} x \left( -\dot{\Phi} \dot{\delta} + \partial_z \Phi \partial_z\delta + \partial_i \Phi \partial^i \delta \right)  \nonumber \\
&\quad +c \int_{t_1}^{t_2} \!\!\!\! \dd t  \int_{\mathbb{R}^{d-1}} \!\!\!\!\!\!\!\! \dd^{d-1} x \left(-\dot{\Phi} \dot{\delta} + \partial_i \Phi \partial^i \delta + m^2_{\rm b} \Phi \delta \right) \nonumber \\
& = \int_{t_1}^{t_2} \!\!\!\! \dd t \int_{\mathbb{R}^+} \!\!\!\! \dd z \int_{\mathbb{R}^{d-1}} \!\!\!\!\!\!\!\! \dd^{d-1} x \, \delta \, \Box_\eta^{(d+1)} \Phi \nonumber \\
&\quad -c \int_{t_1}^{t_2} \!\!\!\! \dd t  \int_{\mathbb{R}^{d-1}} \!\!\!\!\!\!\!\! \dd^{d-1} x \, \delta \left(\Box_\eta^{(d+1)} \Phi - m^2_{\rm b} \Phi + \frac{1}{c} \partial_z \Phi  \right) \, ,
\end{align*}
where, on the right hand side of the last equality above, the integration by parts in $z$ in the bulk action has contributed to the boundary term. 

The extrema of the action, $dS(\Phi) = 0$, are
\begin{equation} \label{variation}
\begin{cases}
\Box_{\eta}^{(d+1)} \Phi = 0 \quad \text{in $\bHo^{d+1}$} \, , \\
\left(\Box_{\eta}^{(d)} - m^2_{\rm b} \right) F = - \dfrac{\rho}{c} \quad \text{in $\bR^d$} \, , \\
\Phi|_{z=0} = F \, , \quad \partial_z \Phi|_{z=0} = \rho \, .
\end{cases}
\end{equation}
The boundary conditions of the problem \eqref{variation} are known as \emph{generalized Wentzell boundary conditions} (WBCs), see \cite{Zahn:2015due} and references therein. These are dynamical boundary conditions, for which there is a boundary field $F$ required to coincide with the restriction of the bulk field at the boundary and to satisfy a Klein-Gordon equation with a source term, which is related to the derivative of the bulk field with respect to the direction orthogonal to the boundary. In the case in which the bulk field is massive, and the field equation is singular at the boundary, the explicit form of these boundary conditions need to be generalized, as the bulk field or its derivative may not be defined at the boundary. We discuss such generalization in Section~\ref{sec:singular}.

\subsection{Regular case}
\label{sec:regular}

A Klein-Gordon field, $\phi$, in PAdS$_{d+1}$, satisfying eq. \eqref{KGPAdS} with $m_0^2 = -(\xi - \frac{d-1}{4d})R$ can be mapped to the problem defined by Eq.~\eqref{KGH} with $m^2 = 0$. This defines, together with appropriate boundary conditions, a regular problem in $\bHo^{d+1}$. We choose the above-introduced WBCs,
\begin{equation} \label{eq:regsystem}
\begin{cases}
\Box_{\eta}^{(d+1)} \Phi = 0 \quad \text{in $\bHo^{d+1}$} \, , \\
\left(\Box_{\eta}^{(d)} - m^2_{\rm b} \right) F = - \dfrac{\rho}{c} \quad \text{in $\bR^d$} \, , \\
\Phi|_{z=0} = F \, , \quad \partial_z \Phi|_{z=0} = \rho \, .
\end{cases}
\end{equation}
Here, the parameter $c$ is taken to be real and we restrict $m^2_{\rm b} \geq 0$, so that $m^2_{\rm b}$ is interpreted as a {\it squared boundary field mass} for the {\it boundary field} $F$. We further assume that the Fourier transforms for $\Phi$, $F$ and $\rho$ exist. It suffices for our purposes to consider these functions to identify tempered distributions.

\subsubsection{Bulk and boundary solutions}

For the bulk field, we take the Fourier transform along the directions orthogonal to $z$,
\begin{equation} \label{eq:Fouriertransf}
\Phi(\underline{x},z) = \int_{\bR^d} \frac{\dd^d\underline{k}}{(2\pi)^{\frac{d}{2}}} \, e^{i\underline{k}\cdot \underline{x}} \, \widehat{\Phi}(\underline{k},z) \, ,
\end{equation}
where $\underline{x} \doteq (t, x_1, \ldots, x_{d-1})$, $\underline{k} \doteq (\omega, k_1, \ldots, k_{d-1})$ and $\widehat{\Phi}$ are solutions of
\begin{equation} \label{eq:regmodeeq}
- \frac{\dd^2}{\dd z^2} \widehat{\Phi}(\underline{k},z) = q^2 \, \widehat{\Phi}(\underline{k},z) \, , \quad
q^2 \doteq \omega^2 - \displaystyle\sum_{i=1}^{d-1} k_i^2 \, .
\end{equation}
We note that the differential operator in the LHS of \eqref{eq:regmodeeq} is of Sturm-Liouville type \cite{Zettl:2005}. Therefore we will work in this framework, whose first step calls for identifying those $\widehat{\Phi}(\underline{k},z)$ which are necessary to construct the fundamental solution associated to \eqref{eq:regmodeeq}. 

For the boundary field and source term, we also take the Fourier transform along all directions,
\begin{subequations} \label{eq:FourierFrho}
\begin{align}
F(\underline{x}) &= \int_{\bR^d} \frac{\dd^d\underline{k}}{(2\pi)^{\frac{d}{2}}} \, e^{i\underline{k}\cdot \underline{x}} \, \widehat{F}(\underline{k}) \, , \\
\rho(\underline{x}) &= \int_{\bR^d} \frac{\dd^d\underline{k}}{(2\pi)^{\frac{d}{2}}} \, e^{i\underline{k}\cdot \underline{x}} \, \widehat{\rho}(\underline{k}) \, .
\end{align}
\end{subequations}

In view of the theory for Sturm-Liouville equations \eqref{eq:regmodeeq} should be treated as an eigenvalue equation on $L^2((0,\infty); \dd z)$ with spectral parameter $q^2$. Being $- \frac{\dd^2}{\dd z^2}$ the standard kinetic operator, its spectrum includes a continuous part, $q^2>0$, with a basis of eigensolutions $\big\{ \widehat{\Phi}_1, \, \widehat{\Phi}_2 \big\}$, with
\begin{align} \label{eq:fundbasisreg}
\widehat{\Phi}_1(\underline{k},z) = \frac{\sin(qz)}{q} \, , \qquad
\widehat{\Phi}_2(\underline{k},z) = - \cos(qz) \, .
\end{align}
Observe that both solutions are square-integrable in a neighbourhood of the origin. We call $\widehat{\Phi}_1$ the \emph{principal solution} at $z=0$, as it is the unique solution (up to scalar multiples) such that $\lim_{z \to 0} \widehat{\Phi}_1(\underline{k},z)/\widehat{\Psi}(\underline{k},z)=0$ for every solution $\widehat{\Psi}(\underline{k},z)$ which is not a scalar multiple of $\widehat{\Phi}_1$. The solution $\widehat{\Phi}_2$ is a nonprincipal solution and is not unique. 

A general solution for $q^2>0$ may then be written as
\begin{equation} \label{eq:reglincomb}
\widehat{\Phi}(\underline{k},z) = A(\underline{k}) \widehat{\Phi}_1(\underline{k},z) + B(\underline{k}) \widehat{\Phi}_2(\underline{k},z) \, .
\end{equation}
From \eqref{eq:regsystem} and \eqref{eq:FourierFrho}, one gets
\begin{equation} \label{eq:ABrhoFreg}
A(\underline{k}) = \widehat{\rho}(\underline{k}) \, , \qquad
B(\underline{k}) = - \widehat{F}(\underline{k}) \, .
\end{equation}
These coefficients depend explicitly in $\underline{k}$, contrarily to the more common Robin boundary conditions. However, it is possible to recover the latter, as explained below.

It remains to obtain the boundary field in terms of the source term. From \eqref{eq:regsystem}, it is easy to obtain
\begin{equation} \label{eq:Fitorho}
\widehat{F}(\underline{k}) = - \frac{\widehat{\rho}(\underline{k})}{c \left[q(\underline{k})^2 - m^2_{\rm b} \right]} \, .
\end{equation}
Hence, the solution for $q^2>0$ may be written as 
\begin{equation} 
\widehat{\Phi}(\underline{k},z) = \rho(\underline{k}) \left[ \widehat{\Phi}_1(\underline{k},z) + \frac{\widehat{\Phi}_2(\underline{k},z)}{c \left[q(\underline{k})^2 - m^2_{\rm b} \right]} \right] \, .
\end{equation}
Observe that the term $q(\underline{k})^2 - m^2_{\rm b}$ contributes to a singular term which corresponds to two simple poles in the Fourier transform. These can be dealt with by means of standard, complex analysis techniques.

\subsubsection{Existence of bound states}

Above, we analyzed the continuous part of the spectrum $q^2>0$ associated with the eigenvalue problem given by \eqref{eq:regmodeeq}. 
Here, we investigate if there exist also negative eigenvalues, $q^2 < 0$, in the point spectrum with eigensolutions which satisfy the WBCs. Contrary to the continuous spectrum, in these case we must look for proper eigenfunctions, that is square-integrable solutions to \eqref{eq:regmodeeq}.

For that let $\lambda = -q^2 > 0$ and consider
\begin{equation}
\widehat{\Phi}_{\rm bs}(\underline{k},z) = - B(\underline{k}) \, e^{-\sqrt{\lambda}z} \, .
\end{equation}
This is certainly a solution of the bulk field equation. If it additionally solves the WBCs for some choice of $\lambda$, it is a \emph{bound state mode solution}, that is, a mode which exponentially decay with $z$.

The WBCs, together with \eqref{eq:Fitorho}, imply that
\begin{equation} \label{eq:lambdaeq}
\sqrt{\lambda} = c \left(q^2 - m^2_{\rm b} \right) = c \left(- \lambda - m^2_{\rm b} \right) \, .
\end{equation}
It is clear that, with $m^2_{\rm b} \geq 0$, if $c\geq 0$ there is no positive $\lambda$ which solves the equation. For $c<0$, the solutions are
\begin{equation}
\lambda = \frac{1}{2c^2} \left(1 - 2 m^2_{\rm b} c^2 \pm \sqrt{1 - 4 m^2_{\rm b} c^2}\right) \, .
\end{equation}
If $m_{\rm b}=0$ there exists \emph{one} strictly positive value of $\lambda$, which corresponds to one negative eigenvalue and, thus, one bound state. If $m_{\rm b}>0$, we have three cases:
\begin{itemize}
\item[(i)] If $c<-1/(2m_{\rm b})$, then there is \emph{no} strictly positive value of $\lambda$, and thus no bound states.
\item[(ii)] If $c=-1/(2m_{\rm b})$, then there is \emph{one} strictly positive value of $\lambda$, which corresponds to one negative eigenvalue and, thus, one bound state.
\item[(iii)] If $-1/(2m_{\rm b})<c<0$, there are always \emph{two} strictly positive values of $\lambda$, corresponding to two negative eigenvalues and, thus, two bound states.
\end{itemize}

\subsubsection{Robin boundary conditions}

It is possible to recover Robin boundary conditions at $z=0$ from the WBCs through a specific choice of the boundary field mass term $m^2_{\rm b} $ and an appropriate limit of the constant $c$.

To see that, choose the boundary field mass such that
\begin{equation}
m^2_{\rm b} = \frac{1}{c \kappa} \, ,
\end{equation}
where $\kappa$ is a real number which is positive for $c>0$ and negative for $c<0$, {\it i.e.}~keeping the squared boundary field mass positive. Then, Robin boundary conditions are recovered in the limit $c \to 0$:
\begin{equation} 
\widehat{\Phi}(\underline{k},z) = \rho(\underline{k}) \left[ \widehat{\Phi}_1(\underline{k},z) - \kappa \, \widehat{\Phi}_2(\underline{k},z) \right] \, .
\end{equation}
The usual Dirichlet boundary conditions, for which $\widehat{\Phi}(\underline{k},0)=0$, correspond to $\kappa = 0$, whereas $\kappa \to \infty$ correspond to Neumann boundary conditions, for which  $\partial_z\widehat{\Phi}(\underline{k},z)|_{z=0}=0$.

In \cite{Dappiaggi:2016fwc} it was found that one bound state mode solution exists when $\kappa < 0$, otherwise no such solution exists. From \eqref{eq:lambdaeq}, if we set $m^2_{\rm b} = \frac{1}{c \kappa}$ and then take $c \to 0^-$, we obtain
\begin{equation}
\sqrt{\lambda} = - \frac{1}{\kappa} \, .
\end{equation}
Hence, if $\kappa < 0$ there is a strictly positive value of $\lambda$, which furthermore agrees with the result in \cite{Dappiaggi:2016fwc}. If $\kappa > 0$, there are no bound states in the limit $c \to 0^+$, also in agreement with the results of \cite{Dappiaggi:2016fwc}.

\subsection{Singular case}
\label{sec:singular}

In the case of a massive scalar field in the Poincar\'e patch of AdS$_{d+1}$, the corresponding field equation for the conformally related field in $\bHo^{d+1}$ is singular at $z=0$ and the previous formulation of the WBCs is no longer valid, as the bulk field or its derivative with respect to $z$ may not be defined at $z=0$. In order to bypass this hurdle, we rewrite the underlying equation of motion in the following way:
\begin{equation} \label{eq:singsystem}
\begin{cases}
\left(\Box_{\eta}^{(d+1)} - \frac{m^2}{z^2} \right) \Phi = 0 \quad \text{in $\bHo^{d+1}$} \, , \\
\left(\Box_{\eta}^{(d)} - m^2_{\rm b} \right) F = - \dfrac{\rho}{c} \quad \text{in $\bR^d$} \, , \\
W_z \left[\Phi, \Phi_1\right] = F \, , \quad W_z \left[\Phi, \Phi_2\right] = \rho \, ,
\end{cases}
\end{equation}
where $\big\{ \Phi_1, \, \Phi_2 \big\}$ is a basis of solutions, $W_z[u,v] \doteq u\frac{\partial v}{\partial z} - \frac{\partial u}{\partial z}v$ is the Wronskian betweens, and where $\Phi_1$ is chosen so that $\widehat{\Phi}_1$ is a principal solution at $z=0$. Since both $\widehat{\Phi}_1$ and $\widehat{\Phi}_2$ turn out to be solutions of an ODE with no first order derivative, see \eqref{eq:singmodeeq} below, their Wronskian is constant in $z$. Hence we can normalize them so that
\begin{equation} \label{eq:condwronskian}
W_z \left[\widehat{\Phi}_1, \, \widehat{\Phi}_2\right] = 1 \, ,
\end{equation}
and $\big\{ \widehat{\Phi}_1, \, \widehat{\Phi}_2 \big\}$ reduce to \eqref{eq:fundbasisreg} in the regular case.

\subsubsection{Bulk and boundary solutions}

Using the same Fourier expansions as in \eqref{eq:Fouriertransf} in \eqref{eq:FourierFrho}, $\widehat{\Phi}$ is now solution of
\begin{equation} \label{eq:singmodeeq}
\left(- \frac{\dd^2}{\dd z^2} + \frac{m^2}{z^2} \right)\widehat{\Phi}(\underline{k},z) = q^2 \, \widehat{\Phi}(\underline{k},z)  \, .
\end{equation}
Here, it is useful to remind ourselves that $m^2 = m_0^2 + (\xi - \frac{d-1}{4d})R$ and to introduce the convenient notation
\begin{equation}
\nu \doteq \frac{1}{2} \sqrt{1+4m^2} \, .
\end{equation}
The BF bound implies that $\nu \geq 0$.

A basis of solutions $\big\{ \widehat{\Phi}_1, \, \widehat{\Phi}_2 \big\}$ satisfying the required properties for the continuous part of the spectrum, $q^2 > 0$, is the following:
\begin{subequations}
\begin{align} 
\widehat{\Phi}_1(\underline{k},z) &= \sqrt{\frac{\pi}{2}} \, q^{-\nu} \sqrt{z} \, J_{\nu}(qz) \, , \label{eq:fundamentalsolutions1} \\
\widehat{\Phi}_2(\underline{k},z) &=
\begin{cases}
    - \sqrt{\dfrac{\pi}{2}} \, q^{\nu} \sqrt{z}  \, J_{-\nu}(qz) \, , & \nu > 0 \, , \\
    - \sqrt{\dfrac{\pi}{2}} \sqrt{z} \left[ Y_{0}(qz) - \dfrac{2}{\pi} \log(q) \right] \, , & \nu = 0 \, .
\end{cases} \label{eq:fundamentalsolutions2}
\end{align}
\label{eq:fundamentalsolutions}%
\end{subequations} 
The solution $\widehat{\Phi}_1$ is the principal solution at $z=0$ and is square-integrable near $z=0$ for all $\nu \geq 0$. The nonprincipal solution $\widehat{\Phi}_2$ is only square-integrable near $z=0$ if $\nu \in [0,1)$, hence, we only apply the WBCs for those values of the mass, namely $m^2 \in [-\frac{1}{4}, \frac{3}{4})$.

Hence, a general solution satisfying WBCs for $q^2 > 0$ and $\nu \in [0,1)$ is
\begin{equation}\label{eq:fundbasissing}
\widehat{\Phi}(\underline{k},z) = A(\underline{k}) \widehat{\Phi}_1(\underline{k},z) + B(\underline{k}) \widehat{\Phi}_2(\underline{k},z) \, ,
\end{equation}
with $A(\underline{k}) = \widehat{\rho}(\underline{k})$ and $B(\underline{k}) = - \widehat{F}(\underline{k})$, where we used \eqref{eq:singsystem} and \eqref{eq:condwronskian}. For $\nu \geq 1$, the only square-integrable solution is given by $\widehat{\Phi}_1$ and no boundary conditions need to be applied at $z=0$.

The boundary field $F$ is still given by the same expression of the regular case,
\begin{equation} \label{eq:Fitorho2}
\widehat{F}(\underline{k}) = - \frac{\widehat{\rho}(\underline{k})}{c \left[q(\underline{k})^2 - m^2_{\rm b} \right]} \, .
\end{equation}
Hence, the bulk field may be written as 
\begin{equation} 
\widehat{\Phi}(\underline{k},z) = \rho(\underline{k}) \left[ \widehat{\Phi}_1(\underline{k},z) + \frac{\widehat{\Phi}_2(\underline{k},z)}{c \left[q(\underline{k})^2 - m^2_{\rm b} \right]} \right] \, .
\end{equation}

\subsubsection{Existence of bound states}

Analogously to the regular case, we investigate if there exists negative eigenvalues, $q^2 < 0$, in the point spectrum of the singular eigenvalue problem given by \eqref{eq:singmodeeq} with proper eigenfuctions which satisfy the WBCs.

Again letting $\lambda = -q^2 > 0$, consider
\begin{equation}
\widehat{\Phi}_{\rm bs}(\underline{k},z) = \sqrt{z} \, K_{\nu}(\sqrt{\lambda}z) \, .
\end{equation}
This is a solution of \eqref{eq:singmodeeq}, and the WBCs, together with \eqref{eq:Fitorho2}, imply that
\begin{equation} \label{eq:lambdanu}
\lambda^{\nu} = c \left(q^2 - m^2_{\rm b} \right) = c \left(- \lambda - m^2_{\rm b} \right) \, .
\end{equation}
For $c \geq 0$ there is no positive $\lambda$ that solves the equation. 

If $c<0$, we have several possibilities. If $m_{\rm b}=0$, then there is one strictly positive root,
\begin{equation}
\lambda = (-c)^{\frac{1}{\nu-1}} \, ,
\end{equation}
corresponding to one bound state. If $m_{\rm b}>0$ and $\nu=0$, there is a positive solution,
\begin{equation}
\lambda = - \frac{1+cm^2_{\rm b}}{c} \, ,
\end{equation}
when $- 1/m^2_{\rm b} < c < 0$, otherwise there is no positive solution, and hence no bound states.
If $m_{\rm b}>0$ and $\nu \in (0,1)$, we cannot find analytical solutions of \eqref{eq:lambdanu}, but we can still obtain the number of positive roots. Let
\begin{equation}
f(\lambda)=\lambda^{\nu}+c\lambda+c m^2_{\rm b} \, .
\end{equation}
We want to know if $f$ has any positive roots for $c<0$ and $\nu \in (0,1)$. First, note that $f(0)=c m^2_{\rm b} < 0$ and that $\lim_{\lambda\to\infty}f(\lambda)=-\infty$ for $\nu \in (0,1)$ and $m_{\rm b} > 0$. Moreover, there is only one maximum at $\lambda_{\rm max} = (-c/\nu)^{1/(\nu-1)}$ with 
\begin{equation}
f(\lambda_{\rm max}) = (1-\nu) \left(\frac{-c}{\nu}\right)^{\frac{\nu}{\nu-1}} + c m^2_{\rm b} \, .
\end{equation}
The maximum is positive, and hence there are two positive roots, if $-\nu^{\nu}(m^2_{\rm b}/(1-\nu))^{\nu-1}<c<0$. Otherwise, if $c=-\nu^{\nu}(m^2_{\rm b}/(1-\nu))^{\nu-1}$, there is one positive root, and if $c<-\nu^{\nu}(m^2_{\rm b}/(1-\nu))^{\nu-1}$ then there are no positive roots.

\begin{figure}[t]
\includegraphics[scale=0.65]{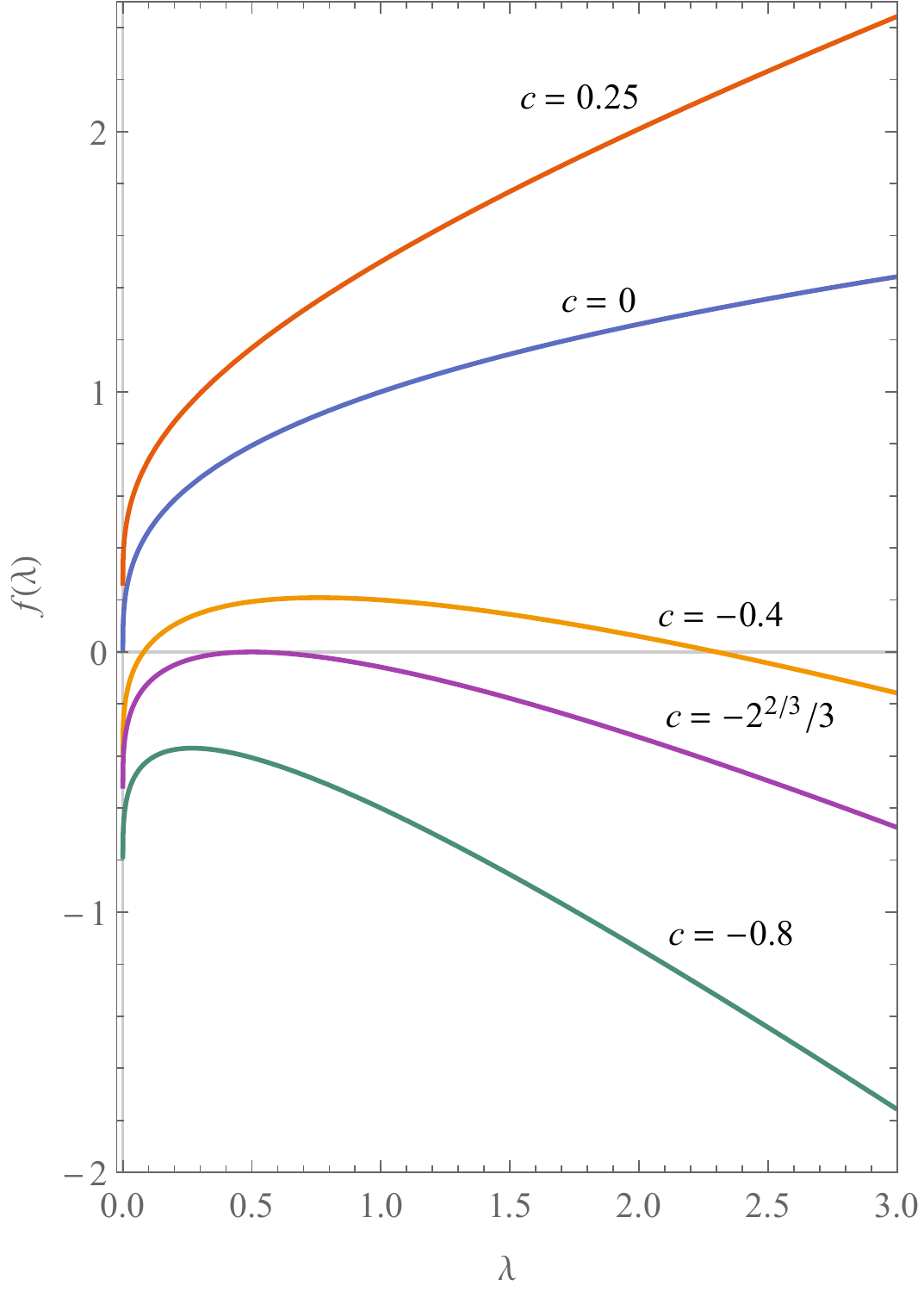}
\caption{\label{fig:plot}Plot of $f(\lambda)=\lambda^{\nu}+c\lambda+c m^2_{\rm b}$ for $\nu=1/3$ and $m_{\rm b}=1$ for different values of $c$.}
\end{figure}

We then conclude that, if $c<0$, $m_{\rm b} > 0$ and $\nu \in (0,1)$:
\begin{itemize}
\item[(i)] If $c<-\nu^{\nu}(m^2_{\rm b}/(1-\nu))^{\nu-1}$, then there is \emph{no} strictly positive value of $\lambda$, and, thus, no bound states.
\item[(ii)] If $c=-\nu^{\nu}(m^2_{\rm b}/(1-\nu))^{\nu-1}$ (or $m_{\rm b}^2 = 0$), then there is \emph{one} strictly positive value of $\lambda$, which corresponds to one negative eigenvalue and, thus, one bound state.
\item[(iii)] If $-\nu^{\nu}(m^2_{\rm b}/(1-\nu))^{\nu-1}<c<0$, there are always \emph{two} strictly positive values of $\lambda$, corresponding to two negative eigenvalues and, thus, two bound states.
\end{itemize}

These results are illustrated in Fig.~\ref{fig:plot}. Note that they are in agreement with the regular case, $\nu=\frac{1}{2}$.

Finally, if we consider the limit in which we recover Robin boundary conditions, by setting $m^2_{\rm b} = \frac{1}{c \kappa}$ and then taking $c \to 0$, we obtain from \eqref{eq:lambdanu} that
\begin{equation}
\lambda^{\nu} = - \frac{1}{\kappa} \, .
\end{equation}
Hence, if $\kappa < 0$ there is a strictly positive value of $\lambda$, and no positive values of $\lambda$ for $\kappa >0$, which agrees with the result in \cite{Dappiaggi:2016fwc}.


\section{Distinguishing structural properties of the generalized Wentzell boundary conditions}
\label{sec:maths}

In this section, we first show that imposing the generalized Wentzell boundary conditions (WBCs) at the PAdS boundary guarantees that the total fluxes of symplectic and energy currents across the boundary vanish, thus showing that the system is closed, as is the case with Robin (hence Dirichlet and Neumann) boundary conditions. We then provide a further explanation so as to why these and the Robin boundary conditions are \emph{distinguished} on account of their interplay the scalar represetation of the isometry group at the conformal boundary.

\subsection{Vanishing symplectic and energy flux across the boundary}

For the bulk field $\Phi$, now assumed to be complex-valued, we may define the \emph{bulk symplectic current} as 
\begin{equation}
J_{\mu} \doteq -i \left(\overline{\Phi} \partial_{\mu} \Phi - \Phi \partial_{\mu} \overline{\Phi} \right) \, .
\end{equation}
It is covariantly conserved, $\partial_{\mu}J^{\mu} = 0$, or equivalently $\dd \ast J = 0$. Using Stokes' theorem,
\begin{align}
0 = \int_{\bHo^{d+1}} \dd \ast J = \int_{\bR^d} \ast J \, ,
\end{align}
which implies that, in combination of \eqref{eq:regsystem} or \eqref{eq:singsystem},
\begin{align}
\int_{\bR^d} \dd^d x \, \left(\rho \overline{F} - \overline{\rho} F \right) = 0 \, .
\end{align}
This is a condition that both the source term $\rho$ and the boundary field $F$ must satisfy. One possibility is that the integrand itself vanishes, which, by \eqref{eq:ABrhoFreg}, implies that the ratio $B/A$ in \eqref{eq:reglincomb} or \eqref{eq:fundbasissing} must be real --- that is, Robin boundary conditions must be imposed at the AdS boundary. But, more generally, the integrand does not need to vanish as long as its integral over the boundary is identically zero. Using \eqref{eq:Fitorho} or \eqref{eq:Fitorho2}, one has
\begin{align*}
{}& \int_{\bR^d} \dd^d x \int_{\bR^d} \dd^d \underline{k}_1 \int_{\bR^d} \dd^d \underline{k}_2 \,
\left[ \frac{\widehat{\rho}(\underline{k}_1) \overline{\widehat{\rho}(\underline{k}_2)}}{c \left[ q(\underline{k}_2)^2 - m_{\rm b}^2 \right]} e^{i(\underline{k}_1-\underline{k}_2)\cdot \underline{x}} \right. \\
&\quad - \left. \frac{\overline{\widehat{\rho}(\underline{k}_1)} \widehat{\rho}(\underline{k}_2)}{c \left[ q(\underline{k}_2)^2 - m_{\rm b}^2 \right]} e^{-i(\underline{k}_1-\underline{k}_2)\cdot \underline{x}} \right] \\
&= \int_{\bR^d} \dd^d \underline{k}_1 \int_{\bR^d} \frac{\dd^d \underline{k}_2}{c \left[ q(\underline{k}_2)^2 - m_{\rm b}^2 \right]} \,
\left[ \widehat{\rho}(\underline{k}_1) \overline{\widehat{\rho}(\underline{k}_2)} - \overline{\widehat{\rho}(\underline{k}_1)} \widehat{\rho}(\underline{k}_2) \right] \\
&\quad \times \delta(\underline{k}_1-\underline{k}_2) \\
&= \int_{\bR^d} \frac{\dd^d \underline{k}}{c \left[ q(\underline{k}_2)^2 - m_{\rm b}^2 \right]} \,
\left[ \widehat{\rho}(\underline{k}) \overline{\widehat{\rho}(\underline{k})} - \overline{\widehat{\rho}(\underline{k})} \widehat{\rho}(\underline{k}) \right] = 0 \, .
\end{align*}
This shows that WBCs guarantees that the total symplectic flux across the boundary vanishes.

Again using \eqref{eq:regsystem} or \eqref{eq:singsystem}, we obtain
\begin{align}
\int_{\bR^d} \dd^d x \, \eta^{\alpha\beta} \partial_{\alpha} \left(\overline{F} \partial_{\beta} F - F \partial_{\beta} \overline{F}\right) = 0 \, ,
\end{align}
which suggests the definition of a \emph{boundary symplectic current}
\begin{equation} \label{eq:boundarysymplecticcurrent}
J^{\partial}_{\alpha} \doteq -i \, c \left(\overline{F} \partial_{\alpha} F - F \partial_{\alpha} \overline{F}\right) \, .
\end{equation}
Note, however, that it is not covariantly conserved, as 
\begin{equation}
\partial^{\alpha} J^{\partial}_{\alpha} = -i \left(\rho \overline{F} - \overline{\rho} F \right) \, ,
\end{equation}
except in the particular case of Robin boundary conditions.

We note that the results presented above for the the symplectic current and for its flux across the boundary apply analogously to the energy density current for a real $\Phi$, defined by $J_{\mu}^E \doteq - T_{\mu\nu} k^{\nu}$, with $k = \partial_t$ and where the \emph{bulk stress-energy tensor} is
\begin{equation}
T_{\mu\nu} = \partial_{\mu} \Phi \partial_{\nu} \Phi - \frac{1}{2} g_{\mu\nu} \! \Big( \partial^{\lambda} \Phi \partial_{\lambda} \Phi + \tilde{m}^2 \Phi^2 \Big) \, ,
\end{equation}
where $\tilde{m}^2 \doteq m_0^2 + \xi R$. Observe that, by using the bulk equations of motion, it holds $\partial^\mu T_{\mu\nu}=0$. We may also define a \emph{boundary stress-energy tensor} as
\begin{equation}
T_{\alpha\beta}^{\partial} = c \Big[\partial_{\alpha} F \partial_{\beta} F - \frac{1}{2} \eta_{\alpha\beta} \! \Big( \partial^{\lambda} F \partial_{\lambda} F + \tilde{m}^2 F^2 \Big) \Big] \, ,
\end{equation}
which is however not covariantly conserved,
\begin{equation}
\partial^{\alpha} T^{\partial}_{\alpha\beta} = \left[c (m_{\rm b}^2 - \tilde{m}^2) - \rho \right] \partial_{\beta} F \, ,
\end{equation}
thus indicating that energy fluxes come from the bulk towards the boundary and leave the boundary into the bulk, as expected on physical grounds.

\subsection{Interplay between the boundary conditions and the boundary isometry group}

Considering boundary conditions of Robin type, but also those as in \eqref{eq:regsystem} or in \eqref{eq:singsystem} might appear at a first glance a mere academic exercise. Yet, as we already discussed in the introduction, the lessons learned from the study of the AdS/CFT correspondence indicate the importance of analyzing the possible interplays between bulk and boundary theories which are both dynamical. In this section, we discuss a different, structural property which indicates that both Robin and Wentzell boundary conditions are distinguished. 

To this avail it is of paramount relevance that the underlying background is static. This allows us to shift from a purely hyperbolic equation such as the one in \eqref{eq:singsystem}, ruled by the wave operator $\Box_\eta^{(d+1)}-\frac{m^2}{z^2}$, to an elliptic problem, governed by
\begin{equation}
K_{\omega,m}\doteq-\nabla^2+\omega^2-\frac{m^2}{z^2} \, ,
\end{equation}
where $\nabla^2=\sum_{i=1}^d \partial_i^2$ and $\omega$ is the Fourier parameter associated to the time coordinate. 

Since we are interested in theories which can be coherently quantized, it is convenient to read the operator $K_{0,m}=-\nabla^2+\frac{m^2}{z^2}$ as the Hamiltonian of the underlying system. From this viewpoint, it is natural to interpret $K_{0,m}$ as a real, symmetric operator, acting on the Hilbert space $L^2(\mathring{\mathbb H}^d)$. Hence, in order for the underlying dynamics to describe a \emph{closed} system, one needs to pick a self-adjoint extension of $K_{0,m}$, whose selection consists in turn on fixing suitable boundary conditions at $z=0$. At the classical level this guarantees that the total flux of symplectic and energy currents across the boundary vanishes, as shown explicitly in the previous section.

To better appreciate our freedom in this choice, we divide the analysis in two cases, $d=1$ and $d>1$. In the first case, and setting without loss of generality $m=0$, the Hamiltonian reduces to the kinetic operator on the half line. By using the theory of deficiency indices \cite[Ch.5]{Moretti:2013cma}, the possible self-adjoint extensions are well-known: they are in one-to-one correspondence with boundary conditions of the form $\Phi|_{z=0}+\tan\alpha\,\partial_z\Phi|_{z=0}=0$, where $\alpha\in [0,\pi)$ can at most be made to be dependent on the spectral parameter, {\it i.e.}~$\alpha=\alpha(\omega)$. This problem has been already investigated in \cite{Dappiaggi:2016fwc}

If $d>1$, the scenario is more intricate since $K_{0,m}$ is either essentially self-adjoint or the associated deficiency indices are infinite. The latter instance occurs for example when $m=0$. In other words there are infinite admissible choices for the ratio between the coefficients $A(\underline{k})$ and $B(\underline{k})$ appearing in \eqref{eq:reglincomb} or in \eqref{eq:fundbasissing}. A physically motivated and mathematically sound selection criterion can be implemented by considering the interplay between the isometry group of the background and the operator $K_{0,m}$. Such problem has been studied only recently in a series of papers \cite{Ibort:2014sua,Ibort2,Pardo}. Another closely related analysis can be found in \cite{Asorey:2015lja,Asorey:2017euv}. We will shortly review them and apply the procedure to the case at hand.

The starting point consists of investigating whether $K_{0,m}$ is an Hermitian operator on the Sobolev space $H^2(\mathbb H^d)$, where, for all $s>0$ and for all integer $d\geq 1$, $H^s(\mathbb{R}^d)=\{\psi\in L^2(M),\;|\;(\mathbb{I}-\nabla^2)^s\psi\in L^2(M)\}$, while  $H^s(\mathbb{H}^d)=\{[\Psi]\;|\Psi\in H^s(\mathbb{R}^d)\;\Psi\sim\Psi^\prime,\;\textrm{iff}\; (\Psi-\Psi^\prime)|_{\mathbb{H}^d}=0\}$ --- see \cite{Ibort2} or \cite{Adams} for a survey of the theory and of the key properties of Sobolev spaces. From now on, we will not write explicitly the symbol of equivalence classes since all our statements do not depend on the representative chosen in each of these classes.

To this avail, we observe that the following Green's formula holds true for all $\Psi,\Psi^\prime\in H^2(\mathbb H^d)$:
\begin{equation}\label{Green_formula}
(\Psi,K_{\omega,m}\Psi^\prime)-(K_{\omega,m}\Psi,\Psi^\prime)=\widetilde{\Sigma}(\Psi,\Psi^\prime) \, ,
\end{equation}
where $(,)$ stands for the inner product in $L^2(\mathbb H^d)$ while 
\begin{equation}\label{boundary_form}
\widetilde{\Sigma}(\Psi,\Psi^\prime)=\langle\Gamma(\Psi),\Gamma(\partial_z\Psi^\prime)\rangle-\langle\Gamma(\partial_z\Psi),\Gamma(\Psi^\prime)\rangle \, .
\end{equation}
where $\langle,\rangle$ is the $L^2$ inner product on the boundary $\mathbb{R}^{d-1}$. At the same time $\Gamma:H^s(\mathbb H^{d})\to H^{s-\frac{1}{2}}(\mathbb{R}^{d-1})$ is the so called \textit{Lions trace} \cite[Chap. 5]{Adams}. For every $s>\frac{1}{2}$ this is a continuous and surjective operator which extends at the level of Hilbert spaces the standard restriction of smooth functions, namely, for every $\Psi\in C^\infty(\mathring{\mathbb H}^{d})\cap H^s(\mathbb H^{d})$, $\Gamma(\Psi)=\Psi|_{z=0}$.  $\widetilde{\Sigma}$ is also known as {\em Lagrange boundary form} (in the case considered in this paper, corresponds to the boundary symplectic current introduced in \eqref{eq:boundarysymplecticcurrent}). A {\em dense} subspace $\mathcal{D}\subseteq\mathcal{H}_{\rm b}\doteq L^2(\mathbb{R}^{d-1})$ is called {\em isotropic} (with respect to $\widetilde{\Sigma}$) if $\widetilde{\Sigma}(\alpha,\beta)=0$ for all $\alpha,\beta\in\mathcal{D}$.

A direct inspection of \eqref{Green_formula} unveils that, for $K_{\omega,m}$ to be a symmetric operator, it is mandatory that $\Sigma$ vanishes on its domain. While this is automatically true if one considers smooth and compactly supported functions on $\mathring{\mathbb H}^d$, from the viewpoint of the boundary Hilbert space, this choice is not informative since $\Gamma[C^\infty_0(\mathring{\mathbb H}^d)]=\{0\}$. Hence it is useful to consider the following relevant sets:
\begin{itemize}
	\item For any $\mathcal{W}\subseteq\mathcal{H}_b\times\mathcal{H}_b$ its {\em $\Sigma$-orthogonal subspace} is 
	\begin{align}
	\mathcal{W}^\perp &\doteq \big\{(\varphi,\varphi^\prime)\in\mathcal{H}_b\times\mathcal{H}_b\;|\;\Sigma((\varphi,\varphi^\prime),(\psi,\psi^\prime))=0, \notag \\
	&\qquad \;\forall (\psi,\psi^\prime)\in\mathcal{W}\times\mathcal{W}\big\} \, ,
	\end{align}
	where $\Sigma$ is the natural generalization of \eqref{boundary_form}, {\it i.e.}
	\begin{equation}
	\Sigma((\varphi,\varphi^\prime),(\psi,\psi^\prime)) = \langle\varphi,\psi^\prime\rangle-\langle\varphi^\prime,\psi\rangle \, .
	\end{equation}
	\item A subspace $\mathcal{W}$ is called {\em $\Sigma$-isotropic} if $\mathcal{W}\subseteq\mathcal{W}^\perp$ and {\em maximally $\Sigma$-isotropic} if $\mathcal{W}=\mathcal{W}^\perp$.
\end{itemize}

The advantage of considering those $\mathcal{W}$ which are maximally $\Sigma$-isotropic is two-fold. On the one hand, since $\Gamma$ is surjective, $\Gamma^{-1}[\mathcal{W}]$ identifies a natural domain of  $K_{0,m}$ on which the right-hand side of \eqref{Green_formula} vanishes automatically. On the other hand, it is possible to give an explicit characterization of these spaces. As a matter of fact, as proven in \cite[Lemma~3.1.4 \& Prop.~3.1.5]{Pardo}, letting $\mathcal{C}:\mathcal{H}_{\rm b}\times\mathcal{H}_{\rm b}\to\mathcal{H}_{\rm b}\times\mathcal{H}_{\rm b}$ be the {\em unitary Cayley transform} 
\begin{equation}\label{eq:Cayley_transform}
\mathcal{C}(\varphi,\varphi^\prime) = \frac{1}{\sqrt{2}}\left(\varphi+i\varphi^\prime,\varphi-i\varphi^\prime\right) \, ,
\end{equation}
it holds that 
\begin{enumerate}
	\item $\mathcal{W}$ is maximally $\Sigma$-isotropic if and only if $\mathcal{W}_c\doteq\mathcal{C}[\mathcal{W}]$ is maximally $\Sigma_c$-isotropic,
	where 
	\begin{equation}\label{Sigma_c}
	\Sigma_c((\varphi,\varphi^\prime),(\psi,\psi^\prime)) = -i(\langle\varphi,\psi\rangle-\langle\varphi^\prime,\psi^\prime\rangle) \, .
	\end{equation}
\end{enumerate}

It follows from the items above that, whenever $\mathcal{W}$ is maximally $\Sigma$-isotropic and for any unitary operator $U$, we can use \eqref{eq:Cayley_transform} to write
\begin{equation}\label{max_isotropic}
\mathcal{W} \doteq \{(\varphi,\varphi^\prime) \in \mathcal{H}_{\rm b}\times\mathcal{H}_{\rm b} \;|\; \varphi-i\varphi^\prime = U(\varphi+i\varphi^\prime)\} \, .
\end{equation}

If we recall that $\Sigma$ generalizes \eqref{boundary_form}, we can identify $\varphi=\Gamma(\Psi)$ and $\varphi^\prime=\Gamma(\partial_z\Psi)$, $\Psi\in H^2(\mathbb H^d)$, which suggests that the choice of any $\mathcal{W}$ as in \eqref{max_isotropic} identifies a specific boundary condition. Formally this can be obtained inverting the identity in \eqref{boundary_form}:
\begin{equation}\label{eq:inverse_formula}
\varphi^\prime=A_U\varphi,\quad A_U\doteq -i(\mathbb{I}-U)(\mathbb{I}+U)^{-1}.
\end{equation}
As observed in \cite{Ibort:2014sua,Ibort2}, for \eqref{eq:inverse_formula} to be both a well-defined mathematical expression and applicable to the case at hand, a sufficient requirement is that two conditions should be met. On the one hand, either $(\mathbb{I}+U)^{-1}$ exists or $-1$ is an element of the spectrum of $U$, which is not an accumulation point. It is noteworthy that, choosing Robin boundary conditions always falls in the first case. We stress that, if we recall the identification $\varphi^\prime=\Gamma(\partial_z\Psi)$, then we also need that $A_U$ is a continuous operator on $H^{\frac{1}{2}}(\mathbb{R}^d)$. Any unitary operator $U:\mathcal{H}_{\rm b}\to\mathcal{H}_{\rm b}$ meeting these requirements will be called {\em admissible}. 

The next step consists of using the structures introduced above to characterize the self-adjoint extensions of the Hamiltonian operator $K_{0,m}$, whenever the deficiency indices are non-vanishing. Within this class, the prototypical case is the one in which we set $m=0$. Hence, from now on we focus our attention on $K\equiv K_{0,0}$, although all results apply also to the other scenarios. 

The first step consists of translating $K$ into an Hermitian quadratic form. Following \cite{Ibort2}, let $U$ be an admissible unitary operator so that $-1$ is not an element of its spectrum. Then we call $Q_U:D_{Q_U}\times D_{Q_U}\subset\mathcal{H}_{\rm b}\times\mathcal{H}_{\rm b}\to\mathbb{C}$, 
\begin{equation}\label{QF}
Q_U(\Phi,\Phi^\prime)\doteq\left( d\Phi,d\Phi^\prime\right)_{\Lambda^1}+\langle\Gamma(\Phi),A_U(\Gamma(\Phi^\prime))\rangle \, ,
\end{equation}
where $(,)_{\Lambda^1}$ stands for the standard $L^2$-pairing between $1$-forms on a Riemannian manifold and where $\Phi,\Phi^\prime\in D_{Q_U}\equiv H^1(\mathbb H^d)$. In \cite{Ibort:2014sua,Ibort2} it has been proven that $Q_U$ enjoys several properties, the most notable being that it is closable, namely there exists a domain $D^\prime_{Q_U}\supseteq D_{Q_U}$ on which $Q_U$ is closed with respect to the norm 
\begin{equation}
\|\Phi\|_Q^2=\|d\Phi\|_{\Lambda^1}^2+(1+C_U)\|\Phi\|^2_{H^1} \, , \quad \forall\Phi\in D^\prime_{Q_U} \, .
\end{equation}

Hence, we can invoke \cite[Th.~2.7 \& 6.7]{Ibort:2014sua} to conclude that the quadratic form $Q_U$ identifies a unique self-adjoint operator $K_U$ such that $D(K_U)= D^\prime_{Q_U}$ and there exists $\chi\in H^2(\mathbb H^d)$ for which $Q_U(\Phi,\Phi^\prime)=(\Phi,\chi)_{H^1}$ for all $\Phi\in D^\prime_{Q_U}$. In this case we set $K_U\Phi^\prime=\chi$  and
\begin{equation}
Q_U(\Phi,\Phi^\prime)=(\Phi,K_U\Phi^\prime) \, , \quad \forall\Phi,\Phi^\prime\in D(K_U) \, .
\end{equation}

In addition, it turns out that $K_U$ is a self-adjoint extension of $K$ uniquely and unambiguously identified by an admissible unitary operator $U:\mathcal{H}_{\rm b}\to\mathcal{H}_{\rm b}$. 

The above digression serves us to recall that the choice of boundary conditions for $K_{\omega,0}$ is strongly tied to the identification of a maximally $\Sigma$-isotropic $\mathcal{W}\subset\mathcal{H}_{\rm b}$, which, in turn, corresponds to selecting a self-adjoint extension for $K$ via an admissible unitary operator $U$. Yet, since the number of the latter is infinite, one might wonder whether it is at least possible to identify a distinguished subclass. 

To this end we observe that the Poincar\'e patch of AdS$_{d+1}$ has isometry group $Iso({\rm PAdS}_{d+1})=O(d-1,1)\ltimes\mathbb{R}^d$, that is the $d$-dimensional Poincar\'e group. On each constant time hypersurface, the relevant subgroup is $E(d-1)\doteq O(d-1)\ltimes\mathbb{R}^{d-1}$. Let $V:E(d-1)\to\mathcal{BL}(L^2(\mathring{\mathbb H}^d))$ be such that 
\begin{equation}
(V(g)\psi)(x)=\psi(g^{-1}x) \, ,\quad\forall \psi\in L^2(\mathring{\mathbb H}^d) \, ,
\end{equation}
where $g^{-1}x$ stands for the geometric action of $g^{-1}$ on the point $x\in \mathring{\mathbb H}^d$. This is a unitary, strongly continuous representation of the Euclidean group. To analyze its interplay with \eqref{QF}, we start by considering $U=\mathbb{I}$. This choice identifies the so-called {\em Neumann quadratic form} $Q_N$ such that
\begin{equation}
Q_N(\Phi)=\|d\Phi\|_{\Lambda^1}^2 \, , \quad D(Q_N)=H^1(\mathbb H^d) \, .
\end{equation}
Observe that \eqref{max_isotropic} yields $\varphi^\prime=0$ if $U=\mathbb{I}$, which, in the case at hand, entails that we are considering Neumann boundary conditions. In addition, a direct calculation shows that $Q_N$ is invariant under the action of $V$, namely, for every $g\in E(d-1)$, it holds $Q_N(V(g)\Phi)=Q_N(\Phi)$. 

Since $E(d-1)$ can also be read as a subgroup of the isometries of the boundary of PAdS$_{d+1}$, one can infer that $V$ has a trace along the boundary (see \cite[Def.~6.9]{Ibort:2014sua}), namely for every $\Phi\in H^1(\mathbb H^d)$, it holds 
\begin{equation}
\Gamma(V(g)\Phi)=v(g)\Gamma(\Phi) \, , \quad\forall g\in E(d-1) \, ,
\end{equation}
where $\Gamma:H^1(\mathbb H^d)\to H^{\frac{1}{2}}(\mathbb{R}^{d-1})$ and $v:E(d-1)\to\mathcal{BL}(L^2(\mathbb R)^{d-1})$ is the strongly continuous, unitary representation implementing the geometric action $v(g)\varphi(y)=\varphi(g^{-1}y)$, $y$ being a point of $\mathbb{R}^{d-1}$. The most notable interplay between traceable representations and self-adjoint extensions of the operator $K$ are a consequence of \cite[Th.~6.10]{Ibort:2014sua}, which entails that $K_U$ is an $E(d-1)-$invariant, self-adjoint extension of $K$ if and only if $[U,v(g)]=$ for all $g\in E(d-1)$. 

From a physical point of view, invariance under the action of the underlying isometry group is a desired property and hence we call {\em distinguished} any self-adjoint extension of $K$ which is $E(d-1)$-invariant. Two examples are certainly of interest to our analysis. In the first we choose $A_U=\cot\alpha \, \mathbb{I}$, $\alpha\in [0,\pi)$, which via Cayley transform corresponds to $U=e^{i\alpha} \, \mathbb{I}$, see \cite[Th.~5.34]{Moretti:2013cma}. We observe that, on the one hand, a multiple of the identity operator $U$ commutes with every representation of $E(d-1)$, while on the other hand, \eqref{max_isotropic}, together with the identification of $\varphi^\prime=\Gamma(\partial_z\Phi)$ and of $\varphi=\Gamma(\Phi)$, yields the standard Robin boundary condition $\varphi^\prime = \cot\alpha \, \varphi$. Hence, Robin boundary conditions identify an $E(d-1)$-invariant self-adjoint extension of $K$.
We should keep in mind that our interest towards the self-adjoint extensions of $K$ arises from having transformed the wave equation on $\PAdS_{d+1}$ (in the massless case) into an eigenvalue problem for the operator $K$. Hence, although $\alpha$ is a constant one might consider to make $\alpha$ dependent on the spectral parameter $\omega$. Yet this option should be discarded since, upon inverse Fourier transform, we would obtain a boundary condition which breaks manifestly  Poincar\'e invariance. 

A second, non trivial self-adjoint operator which commutes with the unitary representation $v$ for $E(d-1)$ is certainly $-\nabla^2_{d-1}$, the (unique self-adjoint extension of the Laplace-Beltrami operator on $\mathbb{R}^{d-1}$. In this case, although the rationale behind our selection criterion is fulfilled, the unitary operator built via Cayley transform from $-\nabla^2_{d-1}$ has a spectrum whose eigenvalues admit $-1$ has an accumulation point. In this case, one is still identifying a self-adjoint extension of $K$, but a more technical analysis is required, using the so-called quasi-boundary triples, see \cite{boundary_triple} for a short review.  

In addition, in view of our need to reinstate the time coordinate via Fourier transform, a more natural choice consists of adding to $-\nabla^2_{d-1}$ a multiple of the identity operator, dependent on the spectral parameter, namely $\nabla^2_{d-1}+(\omega^2+m^2_{\rm b})\mathbb{I}$ where $m^2_{\rm b}> 0$ is a constant. By considering once again \eqref{max_isotropic} together with the identification of $\varphi^\prime=\Gamma(\partial_z\Phi)$ and of $\varphi=\Gamma(\Phi)$, we realize that this choice consists of considering the boundary condition $\varphi^\prime=(-\nabla^2_{d-1}+\omega^2+m^2_{\rm b})\varphi$, which, after an inverse Fourier transform with respect to $\omega$ yields exactly the Wentzell boundary condition \eqref{eq:regsystem} and \eqref{eq:singsystem}. In other words both Robin and Wentzell boundary conditions are distinguished in view of their interplay with the action of the boundary isometry group on the underlying spaces of functions.


\section{Conclusions}
\label{sec:conclusions}

In this paper, we have considered a real, massive scalar field in the Poincar\'e fundamental domain of AdS in $d+1$ dimensions, subject to dynamical boundary conditions of generalized Wentzell type at the PAdS boundary. We solved the full system, for both the bulk and boundary fields, and verified that, depending on the values of the parameters of the theory, there might exist zero, one or at most two bound state mode solutions. Although we have not dwelt into the quantization of the underlying model, this result offers a clear indication concerning those values of the mass and of the curvature coupling parameter for which we can expect or rule out the existence of a ground state. Finally, we analyzed what makes this choice of dynamical boundary conditions distinguished, as they are invariant under the action of the isometry group of the PAdS boundary, and imply zero symplectic and energy density total flux accross the boundary.

As a perspective, we outline that in order to obtain the quantization of the theory, the first step will consist of constructing the bulk propagator / fundamental solutions, relating it to the one which stems from the boundary theory. It is of particular interest to obtain a map from a Hadamard state of the boundary theory to a Hadamard state of the bulk theory, which would constitute an AdS counterpart to the result in asymptotically flat spacetimes \cite{Dappiaggi:2017kka}. This is work in progress.

Finally, we note that the problem that we have studied in this paper belongs to a class of systems, those with dynamical boundary conditions, that are of relevance for a broad spectrum of physical models and theories. The techniques that we have employed in this work are applicable to different problems, ranging from condensed matter to gravitational physics, as well as quantum gravity and high energy physics.


\begin{acknowledgments}
The work of C.~D.\ was supported by the University of Pavia. The work of H.~F.\ was supported by the INFN postdoctoral fellowship ``Geometrical Methods in Quantum Field Theories and Applications'', and in part by a fellowship of the ``Progetto Giovani GNFM 2017 -- Wave propagation on lorentzian manifolds with boundaries and applications to algebraic QFT'' fostered by the National Group of Mathematical Physics (GNFM-INdAM). H.~F.\ also acknowledges the hospitality of the ICN-UNAM and their support through a PAPIIT-UNAM grant IG100316. The work of B.~A.~J.-A.\ was supported by a Consejo Nacional de Ciencia y Tecnolog\'ia (CONACYT, M\'exico) project 101712. B.~A.~J.-A.\ also acknowledges the hospitality of the INFN -- Sezione di Pavia during the realization of part of this work, as well as the support of an International Mobility Award granted by the Red Tem\'atica de F\'isica de Altas Energ\'ias (Red FAE-CONACYT).
\end{acknowledgments}


\end{document}